 \documentclass[12pt, draftclsnofoot, onecolumn]{IEEEtran}
\IEEEoverridecommandlockouts %showing the thanks at the left bottom of the first page
\usepackage{epsfig,latexsym}
\usepackage{float}
\usepackage{indentfirst}
\usepackage{amsmath}
\usepackage{amssymb}
\usepackage{times}
\usepackage{graphicx}
\usepackage{epstopdf}
\usepackage{cancel}
\usepackage{epstopdf}
\usepackage{ulem}
\usepackage{xcolor}
\begin{document}
\title{How Much Communication Resource is Needed to Run a Wireless Blockchain Network?}
\author{\IEEEauthorblockN{Lei Zhang, Hao Xu, Oluwakayode Onireti, Muhammad Ali Imran and Bin Cao} 
	\thanks{Lei Zhang (Lei.Zhang@glasgow.ac.uk), Hao Xu, Oluwakayode Oniretiand and Muhammad Ali Imran are with the James Watt School of Engineering, University of Glasgow, Glasgow, G12 8QQ, UK. Bin Cao is with Beijing University of Posts and Telecommunications, China.} }
%
%\vspace{-0.3in}
%
%
%
%\vspace{-0.2in}
\maketitle 
\thispagestyle{empty}
\begin{abstract}

 Blockchain is built on a peer-to-peer network that relies on frequent communications among the distributively located nodes. In particular, the consensus mechanisms (CMs), which play a pivotal role in blockchain, are communication resource-demanding and largely determines blockchain security bound and other key performance metrics such as transaction throughput, latency and scalability. Most blockchain systems are designed in a stable wired communication network running in advanced devices under the assumption of sufficient communication resource provision. However, it is envisioned that the majority of the blockchain node peers will be connected through the wireless network in the future. Constrained by the highly dynamic wireless channel and scarce frequency spectrum, communication can significantly affect blockchain's key performance metrics. Hence, in this paper, we present wireless blockchain networks (WBN) under various commonly used CMs and we answer the question of how much communication resource is needed to run such a network. We first present the role of communication in the four stages of the blockchain procedure. We then discuss the relationship between the communication resource provision and the WBNs performance, for three of the most used blockchain CMs namely, Proof-of-Work (PoW), practical Byzantine Fault Tolerant (PBFT) and Raft. Finally, we provide analytical and simulated results to show the impact of the communication resource provision on blockchain performance.

\end{abstract}
%\IEEEpeerreviewmaketitle
%\setlength{\baselineskip}{1\baselineskip}
%\newtheorem{definition}{Definition}
%\newtheorem{fact}{Fact}
%\newtheorem{assumption}{Assumption}
%\newtheorem{theorem}{Theorem}
%\newtheorem{lemma}{Lemma}
%\newtheorem{corollary}{Corollary}
%%\newtheorem{proposition}{\underline{Proposition}}[section]
%\newtheorem{proposition}{Proposition}
%\newtheorem{example}{Example}
%\newtheorem{remark}{Remark}
%%\newtheorem{proof}{\underline{Proof}}[section]
%\newtheorem{algorithm}{Algorithm}
\newcommand{\quickwordcount}[1]{%
  \immediate\write18{texcount -1 -sum -merge -q #1.tex output.bbl > #1-words.sum }%
  \input{#1-words.sum} words%
}

\section{Introduction}
Blockchain, well recognized as the backbone technology of Bitcoin, has become a revolutionary data management framework through establishing consensuses and agreements in a trust-less and distributed environment.
It offers an immutable, transparent, secure and auditable ledger in a trust-less distributed environment, to verify the integrity and traceability of information/assets during their life cycle. In addition, without a central authority's involvement, blockchain-enabled smart contracts can significantly reduce manual interventions and thus improve efficiency. 
%Blockchain protocol structures information in a chain of blocks, where each block records transactions\footnote{Transactions are generally regarded as any valuable information exchange between interested parties.} in a manner that the records are 100\% secure and cannot be altered by any malicious/unauthorized users unless they possess resource that is more than the security bound. 
It thus has shown great potentials in various fields such as financial services, energy trading, supply chain, identity management, and the Internet of Things (IoT) \cite{Cao2019}. As such, Gartner forecasts that blockchain will generate an annual business value of more than US \$3 trillion by 2030 \cite{Granetto2017}.
\subsection{Background and Related Work}
The consensus mechanism (CM, aka. consensus algorithm or consensus protocol), which ensures an unambiguous ordering of transactions and guarantees the integrity and consistency of blockchain across geographically distributed nodes, plays a key role in blockchain. CM largely determines blockchain system security bound and performance such as transaction throughput, delay, and node scalability. Depending on application scenarios and performance requirements, different CMs can be used. In a permissionless public chain, nodes are allowed to join/leave the network without permission and authentication. Therefore proof-based algorithms (PoX) such as Proof-of-Work (PoW) \cite{nakamoto2008bitcoin}, Proof-of-Stake (PoS) \cite{vasin2014blackcoin} and their variants are commonly used in many public blockchain applications (e.g., Bitcoin, Ethereum). PoX algorithms are designed with excellent node scalability performance through nodes competition. However, they could be very resource-demanding. For instance, recent study estimates of Bitcoin's electricity consumption range between 0.1\% - 0.3\% of global electricity use \cite{bendiksen2018bitcoin}. Also, these CMs have other limitations such as long transaction confirmation latency and low throughput. 

Unlike the public chain, the private and consortium blockchains prefer to adopt lighter protocols such as Raft \cite{ongaro2014search} and practical Byzantine Fault Tolerance (PBFT) to reduce computational power demand and improve the transaction throughput. This property is critically important to the application scenarios of blockchain-enabled IoT ecosystems, which are typically composed of low-cost and low-power devices. Raft, which is used by the private chain, does not protect the integrity of transactions from malicious attacks, but enables the Crash Fault Tolerance (CFT) for the applying system \cite{ongaro2014search}. To protect the system from malicious users, PBFT was proposed in \cite{Castro1999} as an improved and practical protocol based on original BFT. A well-known example of PBFT implementation is the HyperLedger Fabric \cite{HyTP-toolbox-web}, part of HyperLedger business blockchain frameworks, which has been adopted by tech giants like IBM and Wall Street Fintech, such as J.P. Morgan \cite{TP-toolbox-web}.

\begin{figure*}
	\centering
	\includegraphics[scale=0.6]{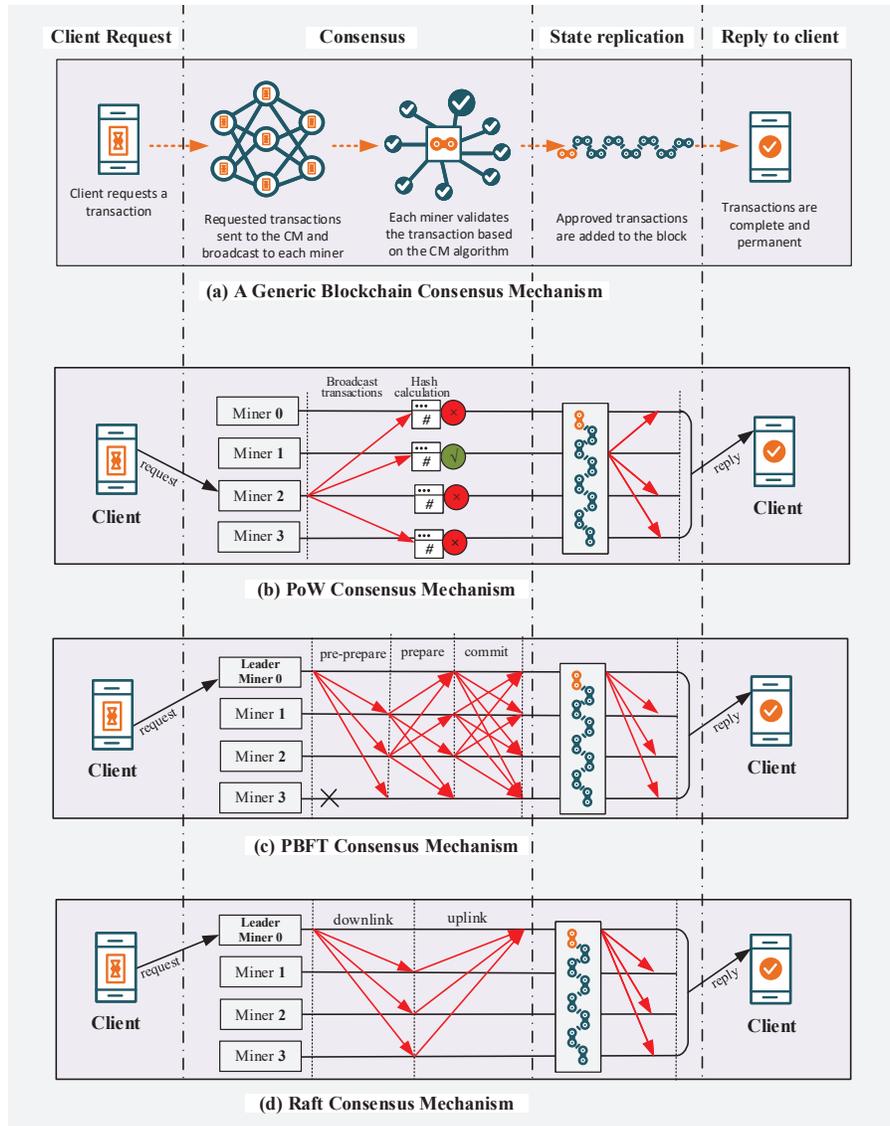}
	\caption{Procedures of three typical CMs (Note that leader re-selection and view change are not included in this figure).  }
	\label{fig:s31}
\end{figure*}

Essentially, blockchain is built on a peer-to-peer network that relies on the frequent communications within the nodes in all four stages of the blockchain procedure, i.e., client request, consensus, state replication and reply to the client, as shown in Fig. \ref{fig:s31}. Among them, consensus requires heavy communication resources and the required amount of communication resources varies among different CMs. For instance, voting based PBFT much more heavily relies on inter-node communications to achieve an agreement among the nodes before it can record the transactions into the blockchain securely. However, the most state of the art blockchains are primarily designed in stable wired communication networks running in advanced devices with sufficient communication resource provision, which refers to allocated communication resource in terms of spectrum, transmission power, receiver sensitivity, number of communication nodes etc. Hence, the performance and security level degradation caused by communication is negligible. Nevertheless, this is not the case for the highly dynamic wireless connected digital society that is mainly composed of massive wireless devices encompassing finance, supply chain, healthcare, transportation and energy.  Especially through the upcoming 5G network, the majority of valuable information exchange may be through a wireless medium.
%\textcolor{blue} {As such, according to the IBM report [6], to have a smart, secure and efficient future, blockchain services will be deployed primarily on hundreds of billions of IoT devices by 2025 and the majority of them will be connected via wireless communications \cite{IBM2015}. For instance, it has been claimed that blockchain used in the connected autonomous driving system can facilitate vehicle safety, data security, data sharing, telematics services, insurance services and thus benefits all involved stakeholders. In addition, blockchain was proposed as an enabler for privacy-preserving contact tracing mobile App to combat COVID-19 \cite{Beeptrace}.} -- Consider to delete, too long now. 
\subsection{Motivations and Contributions}
When combining blockchain with communication (especially wireless communication), the majority of work is focused on how to use blockchain to facilitate the communication resource trading, sharing and optimization (i.e., blockchain for communication) \cite{qiu2019blockchain}\cite{liu2018distributed}. However, from another equally or more important and fundamental aspect, there are few papers systematically analyzed how communication plays a role in blockchain and optimize blockchain performance (i.e., communication for blockchain), which is the main motivation of the paper.

More technically, the frequent communications among nodes can increase the difficulty of harming agreements by the malicious nodes. Thus, any communication link failure between two peer nodes in wireless blockchain may degrade the security level of the blockchain network. From this point of view, it is not a surprise that providing more frequency spectrum or more transmission power can improve the communication link quality and result in a better blockchain security level. However, this is at the cost of power and spectrum efficiency. To guide the deployment and application of blockchain in the wireless environments, the following questions are to be answered. 
\begin{itemize}
	\item  To the best of the authors' knowledge, a framework that presents the communication procedure and roles in the wireless blockchain networks (WBN) is missing. 
	% maybe use WBN to replace wireless blockchain system
	\item  It is not clear how communication can affect blockchain performance such as scalability, throughput, latency and security level.
	\item It is worth to investigate how much communication resource is needed to run a WBN securely.
	\end{itemize}
	
	Recognizing the problems and challenges, the contribution of the paper is summarized as follows. 
\begin{itemize}	
	\item We first present WBN under various commonly used CMs (e.g., PoW, PBFT, Raft), with different network typologies (e.g. mesh, tree, etc.) and communication protocols (e.g. grant-based or contention-based). 
	
	\item We extract and analyze the communications related to four steps of blockchain, i.e., client request, consensus, state replication and reply to the client for PoW, PBFT and Raft based blockchains, respectively. The analysis and model establish a solid foundation to further analyze and optimize the WBN performance.
	\item We bridge the analytical relationship between blockchain performance (in terms of scalability, throughput and latency, and security) and communication resource provision (in terms of spectrum, transmission power and receiver sensitivity) for the three CMs in different communication protocols, to provide a valid tool to estimate how communication resource provision can affect the WBN performance. Due to the openness of wireless channel, interference and intentional radio jamming is also considered in our system. 
	\item We provide extensive analytical and numerical results to show how communication resource provision can affect WBN performance. 
\end{itemize}

\section{Blockchain Performance Metrics}

Security bound, node scalability, transaction throughput and latency are the four most important metrics to measure the blockchain performance. These metrics are largely determined by the blockchain data structure design and CM selection, although those metrics are contradictory to each other to some degree. For instance, in Bitcoin, transactions packed in each block can be confirmed only if six or more blocks are generated afterwards. This protocol design is to prevent the double-spending issues (thus maximize the security performance), which can significantly degrade the transaction throughput and latency performance. From the CM perspective, each CM has its unique privileges and drawbacks, which makes it a tangled choice in real-world applications in order to balance the needs of different prospects. The performance comparison of the blockchain consensuses is summarized in Table I. 
\begin{table*}
	\centering
	\caption{Performance comparison of commonly used blockchain consensuses (Note that transaction latency is reciprocal to transaction throughput, thus it is not listed in the Table.)}
	\begin{tabular}{|c|c|c|c|c|c|c|c|}
		\hline
 \begin{tabular}[c]{@{}l@{}}Consensus\\   Mechanism  \end{tabular}   & \begin{tabular}[c]{@{}l@{}}Suitable Type\\   of Blockchain \end{tabular} & \begin{tabular}[c]{@{}l@{}}Transaction\\   Throughput\end{tabular} & Scalability & \begin{tabular}[c]{@{}l@{}}Security\\   Bound\end{tabular}  & \begin{tabular}[c]{@{}l@{}}Communication\\   Complexity\end{tabular} & \begin{tabular}[c]{@{}l@{}}Spectrum\\   Requirement\end{tabular} &	\begin{tabular}[c]{@{}l@{}}Representative\\   Project \end{tabular}\\ \hline
	PBFT  \cite{Castro1999}      & \begin{tabular}[c]{@{}l@{}}Private/\\   Consortium\end{tabular}   & High                                                               & Low         & $3f+1$         & $2N^2+N$                                              & $2N+1$  &\begin{tabular}[c]{@{}l@{}}HyperLedger\\   Fabric v0.6\end{tabular} 	                                                            \\ \hline
	Raft \cite{ongaro2014search}        & Private & Very High                                                          & Medium      & $2f+1$        & $2N$                                                                   & $N+1$  &Quorum 2.0 	                                                               \\ \hline
	PoW \cite{nakamoto2008bitcoin}      & Public & Low                                                                & High        & $2f+1$           & $2N$                                                                   & $2$   &\begin{tabular}[c]{@{}l@{}}Ethereum\\   Geth\end{tabular}	                                                                \\ \hline
	\end{tabular}
\end{table*}

\paragraph{Security Bound} it is the lifeline of blockchain technology as the security must be guaranteed to validate the transactions stored in blocks.  Security bound can be defined as the maximum faulty or byzantine nodes $f$  supported/tolerated by the consensus protocol. Hence, CMs provide strategies of defense against in-activities and byzantine attacks with security bounds. %, which is defined by either the capacity of the whole network for proof-based CMs or the number of faulty nodes in the whole network for vote-based CMs. 
Typical security bound for PoW is considered as $2f+1$, which means the blockchain will compromise if more than 50\% of the network's resource capacity is possessed by a single party, under perfect communication and non-interruptive service. Differently, the voting-based CMs define the number of faulty nodes as either inactive or malicious, which sends misinformation to imperil the whole network. Under the assumption of perfect communications, generic PBFT allows 1/3 of overall nodes are either byzantine (i.e., malicious user) or faulty, and Raft gives a fair performance with 50\% fault tolerance capability but can not tolerate any malicious node. 

\paragraph{Node scalability} this is a metric to measure the capacity of the system to handle the increasing number of nodes. As shown in Table I, proof-based CMs are designed with excellent node scalability performance through nodes competition. In theory, PoW can hold as many users within the networks without considering the communication burden. However, in practice, considering that all transactions and mining results should be broadcasted and received by all nodes, the spectrum demand in WBNs can be unaffordable when the network is extremely large. When it comes to the voting-based CMs, for instance, PBFT relies on heavy inter-node communications. As the size of the node number grows, the required communication resource provision increases rapidly, resulting in low efficiency and poor scalability. Thus, from the communication resource provision perspective, the PBFT-based blockchain hardly scales up to 100 nodes [14]. 

\paragraph{Transaction throughput and latency} they are two important but reciprocal performance metrics. Transaction throughput is measured by transaction per second (TPS) and transaction latency describes the time duration from transaction request to confirmation. In general, the proof-based consensus suffers from low throughput, due to its time guarded characteristics. On the other hand, a vote-based CM has better liveness, and it can conclude the consensus in a rapid manner; hence it yields greater throughput. For instance, the TPS is normally limited to 7 in Bitcoin and 20 to 30 in Ethereum. The transaction confirmation delay is typically as large as 60 minutes in Bitcoin and 3 minutes in Ethereum [5]. On the other hand, a voting-based blockchain network can achieve a transaction throughput in the range of 100 to 1000 TPS with the current physical communication limits. Note that the communication throughput can be a bottleneck to transaction throughput since a large amount of message exchanges are required for consensus achievement. Hence, transaction throughput and latency are also dependent on the number of nodes in the consensus network.

\section{Roles and Impacts of Communication to Blockchain }
Communication is the means to achieve a consensus in distributed trustless systems. Despite various CM, communication plays an important role in making sure that the CM can achieve its goals of integrity, security and latency. 
\subsection{Communication roles and procedures}
For any type of blockchain network, we can split the whole blockchain process into client request, consensus, state replication and reply to clients, as shown in Fig. \ref{fig:s31}. The first, third and fourth stages in three considered CMs are the same. However, the consensus procedures in different CMs are significantly different. 

\paragraph{Client request} a blockchain transaction is triggered by a client request, where the client has a demand for recording a transaction into the blockchain, as shown in Fig. \ref{fig:s31} (a). More concretely, for the PoW based CMs shown in Fig. \ref{fig:s31} (b), the client will first broadcast the transaction to the consensus network consisting of a number of miners\footnote{Note that miner is used in PoW based blockchains refers to the verification node. Thus, it is exchangeable with the consensus node in this article.} to make sure at least one miner successfully receives this request.
Unlike the PoW, in PBFT/Raft, the transaction from the client will be sent to an appointed leader (though it can be replaced by the view change procedure \cite{Castro1999}) of the wireless blockchain consensus network through unicast communication (though broadcasting can be used, but only the appointed leader can process the broadcast information). From this point of view, the client request has light communications resource demand for all considered CMs. 

\paragraph{Consensus} after the transaction request is submitted to the consensus network\footnote{Note that in most of the blockchain protocols, unconfirmed transactions will be first put into a transaction pool and will be picked up and validated by consensus network based on a certain policy.}, the consensus process will determine the record. From the communication perspective, the consensus is the most communication demanding procedure for all CMs. As for PoW shown in Fig. \ref{fig:s31} (b), any miner(s) who received the client request will be responsible for broadcasting the transaction to the whole consensus network (miner 2 in Fig. \ref{fig:s31}), and ideally, all miners are aware of this new transaction after this procedure. Depending on the network size and communication protocols, the communication complexity in this stage can vary significantly. 
For example, in a global blockchain such as Bitcoin, all miners located in different locations over the world should receive the transaction information within a limited time frame. The communication protocol is TCP/IP based; thus, a significant number of relays/routers are required. \textcolor{red}{When} the coverage is in a small local area network (LAN), all nodes could have sufficient power to cover all other nodes; thus, no relay/routing is required. The consensus in PoW is through a computational competition that involves calculating a predefined-difficulty hash value, which is communication free processing. 

For PBFT, as shown in Fig. \ref{fig:s31} (c), the communication procedure is quite different from PoW/PoS, where the security of the system is achieved from the high computational competition, PBFT has low computational complexity, and the security relies on the frequent inter-node communications. As it can be seen from Fig. \ref{fig:s31} (c), three stages are required to achieve an agreement among the nodes. Within each stage, broadcast communications are needed to exchange the information among nodes. From this point of view, PBFT is a very communication resource-demanding protocol and the scalability of the PBFT is poor. Thus, PBFT is typically used in consortium or private chains consisted of a limited number of nodes. 

Raft also relies on the inter-node information exchange to achieve a consensus among the nodes, as shown in Fig. \ref{fig:s31} (d). It can be seen that Raft has two stages and the leader (i.e., node/miner 0 in the Fig.\ref{fig:s31}) will first broadcast the transaction to all followers by downlink (DL) transmission, upon the successful reception of the transaction, followers will verify the transaction and reply to the leader in the uplink (UL) transmission. Obviously, the required communication complexity is less than PBFT. 

\paragraph{State replication} once the consensus has been reached, successfully verified transactions will be written into a new block by the leader (or winning miner in PoW) and connected with the existing longest blockchain through the hash value. It is important that all nodes in the network can have the latest result from the leader node. More specifically, for PoW, the winning miner will broadcast such information to inform other miners while the leader of PBFT/Raft will perform the same role to other replicas. Blockchain transactions are verifiable by checking the hash tree. Though the communication resource provision can be lighter compared to consensus, however, from each miner's perspective, synchronization delay is one important metric since a longer replication delay means outdated chain information, thus the miner will be less competitive in the next round mining.    
\paragraph{Reply to client}
for all CMs, the client can request the consensus network to confirm whether a transaction is in the blockchain. Blockchain transactions are verifiable by checking the hash tree. The communication complexity can be as low as the client request.
\subsection{Communications impact blockchain performance metrics}
%\subsection{\sout{How} Communications impact blockchain performance metrics  }
In this subsection, we will qualitatively analyze how communication can affect the performance metrics of blockchain in a wireless environment, and the numerical results are given in Sections IV and V. 
\paragraph{Security level} communication reliability is critically important for blockchain security level, especially for voting based CMs. For example, in a wireless LAN Raft, the followers (all other nodes other than the leader) use the multi-access UL to confirm the transaction to the leader (Fig. \ref{fig:s31} (d)). Given fewer communication resources, it is more likely that the collision among users (contention-based communication system) or longer queuing time (in a grant-based communication system) can happen in multi-access, thus a longer time is needed to finish all required confirmations. However, with a time out setting, the network has to decide within a time frame no matter the number of nodes feedback required. Thus, less communication resource provision leads to fewer nodes' confirmations, which will potentially lower the security level of the network since it requires at least 50\% of node confirmations. Power is another domain that can affect the security level. A higher Tx power can achieve more extensive wireless coverage. Thus, more nodes are informed, and potentially more nodes can confirm the transactions, resulting in a higher security level. 
A typical WBN faces security threats from compromised nodes and communication failure. In the worst case, the malicious node will not only send false information but also jams the network via spectrum jamming and spoofing. Such an attack will reduce the security of the WBN. 
\paragraph{Scalability} communication resource provision is a significant factor that determines the blockchain scalability. It is well known that the scalability of PBFT is limited, and thus they can only be adopted in small networks. In wireless consensus networks, the scalability can as well be affected by the transmission (Tx) power since lower power leads to a smaller coverage and thus smaller scalability when the node density is fixed. However, it is worth knowing that PoW based systems scalability is not sensitive to the communication resource provision as much as PBFT based systems. However, Tx power can be a bottleneck for the large scale PoW due to the coverage issues. 
\paragraph{Throughput and latency} throughput and latency are reciprocal, and they are jointly decided by blockchain protocol and communication resource provision. For example, in the PBFT system, more communication resource will make each phase to complete quicker, and thus, the transaction throughput will be enhanced. Larger available communication resources such as spectrum, bandwidth, and power can lead to a low latency blockchain. Tx power and communication protocols can also be the determining factors for the throughput and latency of the WBN. Higher Tx power will lead to broader coverage; thus, less or no relays/routers will be needed, which leads to a reduction in the communication time. The network size also determines the number of transactions that can be recorded to the blockchain. 
\section{How much communication resource is needed? } 
In Section III, we have discussed the communication procedures of different CMs and qualitatively analyzed how communication resource provision can affect wireless blockchain performance and security. In this section, we will demonstrate the communication impacts on the blockchain by using analytical and numerical results. 
\subsection{Communication complexity and spectrum requirement}
\paragraph{Communication complexity} among the CMs, PBFT has the largest communication demand since the mechanism relies on inter-node communication to avoid the interruption from malicious users. \textcolor{red}{We} first consider the communication complexity which refers to the number of communications between the transmitters and the receivers. Given the total number of $N$ nodes in the consensus network, from Fig. \ref{fig:s31} (c), PBFT needs a total of $N+2N^2$ communication at the three stages since all nodes have to communicate to all other nodes in the network. For Raft, the communication is to confirm the votes from at least 50\% of the nodes and the communication complexity for both uplink (from followers to head) and downlink (from head to followers) is up to $2N$. For PoW, the client request should be first broadcasted to all other nodes (miners), then the first miner to successfully calculate the hash value also broadcasts the results to all other nodes. Hence, from this point of view, the communication complexity of the PoW is the same as the RAFT. The communication complexity of the three CMs is summarized in Table I.  
\paragraph {Spectrum requirement} it refers to the required communication spectrum resource in the WBN. Unlike the communication complexity that measures the number of receiver processes, spectrum requirement is equal to the number of transmitter processes (proportional to the required number of resource blocks/transmission slots). Noting that the required  spectrum in WBN is dependent on network topology, in the rest of Sec IV, we consider the simplest case where all nodes are covered by others' radio power, without requiring a relay. While the generic case will be discussed in Sec V. Given this assumption, the overall spectrum requirements are $2N+1$, $N+1$ and $2$ for PBFT, RAFT and PoW respectively. 

\begin{figure}[tbhp]
\includegraphics[width =\textwidth]{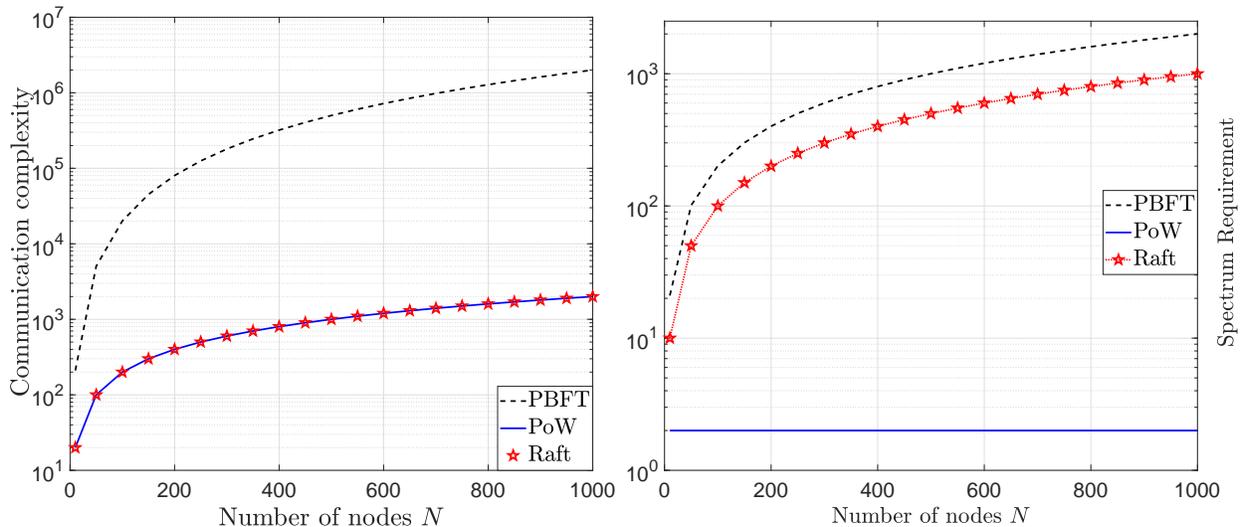}
\caption{The communication complexity (left) and spectrum requirement (right) of PBFT, RAFT and PoW}
\label{fig:s41}
\end{figure}

Fig. \ref{fig:s41} shows the computational complexity and spectrum requirement of PoW, Raft and PBFT CMs. It can be seen that PBFT can have much higher communication complexity than RAFT and PoW in Fig. \ref{fig:s41} (a). Furthermore, Fig. \ref{fig:s41} (b) shows that the spectrum requirement does not change with the number of nodes in PoW while PBFT and Raft required much more spectrum resources than PoW.

\subsection{Tx power}
\begin{figure}[tbhp]
\centering
	\includegraphics[scale = 0.55]{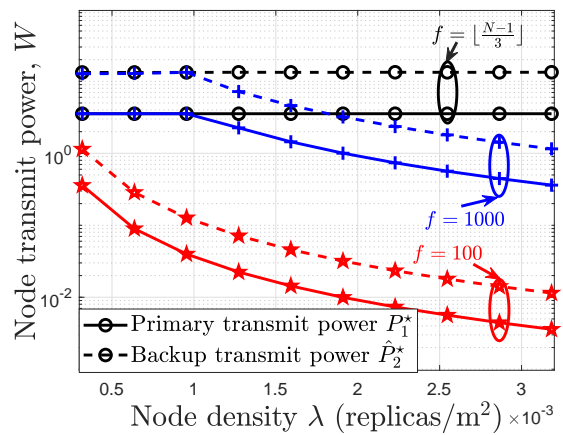}
	\caption{Tx power required for viable wireless PBFT consensus with $f$ faulty nodes.}
		\label{fig:s42}
\end{figure}

In this numerical result, we consider a blockchain-enabled IoT ecosystem composed of battery-powered devices with PBFT as the CM. In a WBN, larger Tx power means a larger coverage area and thus more nodes can receive the consensus information exchange, resulting in a high-security level. The viable area thus defines the minimum WBN's coverage area that meets all the constraints required for a successful implementation of the CMs \cite{Onireti2019}. The viable area is decided by parameters such as the node transmit power and receiver (Rx) sensitivity, channel propagation parameters, number of faulty nodes $f$, transmission interval $v$, and the end-to-end success probability requirement. The viable area ensures that the minimum number of nodes is activated in each view of the wireless PBFT network for a fixed number of faulty nodes $f$, fixed transmission interval $v$ and end-to-end success probability $P_s$, thus leading to significant energy savings and performance improvement. In our simulation set-up, the IoT devices (nodes) are uniformly distributed within the wireless PBFT coverage area with density $\lambda$. The primary node (leader) is located at the origin and we limit its coverage radius to $1000~\mathrm{m}$. 
%It is important to note that the wireless PBFT network coverage area is limited by the primary's coverage, and in particular, the primary's transmit power, since all nodes within the primary's coverage receive the pre-prepare message, which initiates the prepare and then the commit phase. 
The Rx sensitivity $\beta=-84.5~\mathrm{dBm}$, the pathloss exponents $\gamma = 4$. Fig. \ref{fig:s42} plots the transmit power required by the primary node ($P_1^\star$) and other nodes ($P_2^\star$) to have a viable wireless PBFT network, for number of faulty nodes $f=100, 1000$, and $ f=\lfloor\frac{N-1}{3}\rfloor$. The header node and replica nodes must transmit with at least $P_1^\star$ and $P_ 2^\star$, respectively, for a successful wireless PBFT operation. As can be seen, the transmit power required to have a viable wireless PBFT network is dependent on the number of nodes (expressed in terms of node density $\lambda$) and the number of faulty nodes $f$.
\subsection{Receiver sensitivity in Raft CM }
\begin{figure*}
\centering
	\includegraphics[width = 0.8\textwidth]{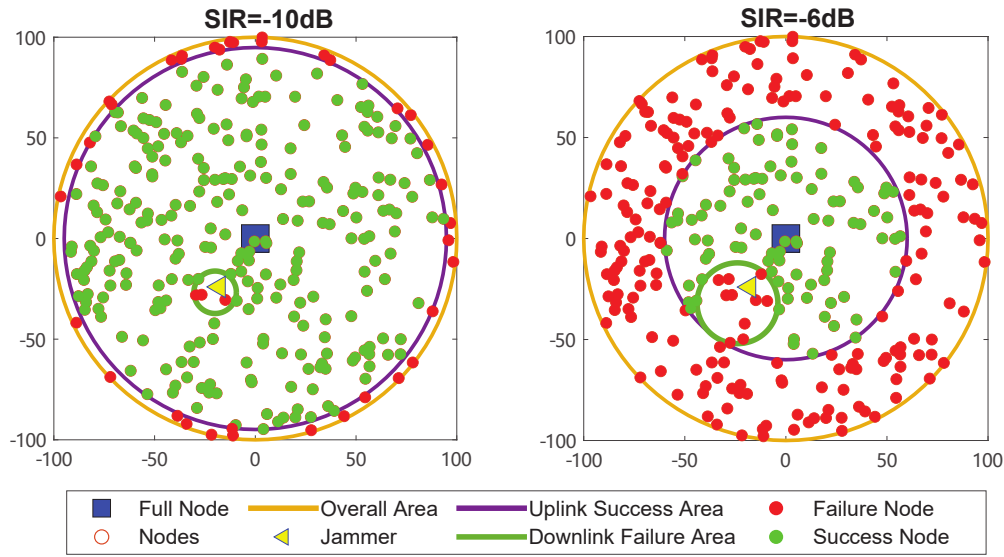}
	\caption{ Raft CM performance under different Rx sensitivities in the presence of a jammer (Red and Green dots refer to successful and unsuccessful votes, respectively).}
		\label{fig:s43}
\end{figure*}

In this illustrative numerical result in Fig. \ref{fig:s43}, we show how Raft CM can be affected by the Rx sensitivity in the presence of a jamming signal. We assume that the equal transmission power and omnidirectional gain are used by all transmitting nodes and the jammer, and the jammer continuously transmits at the same frequency bands used by other nodes thus generating interference to the consensus network. The leader is set in the origin of the coverage circle and all other nodes are uniformly randomly distributed within the circle. The number of nodes $N=300$ in the radius $R=100$m, and the pathloss exponents $\gamma$ is assumed as 2.5 for both UL and DL. To measure Rx sensitivity, we assume a receiver can detect the information and confirm the transaction if the Rx signal-to-interference ratio (SIR) is larger than the threshold.

In Fig. 4, green nodes are successful on both UL and DL, while the red nodes failed either in UL or DL transmission due to their received SIR being lower than the threshold. A higher sensitivity (SIR $= -10$ dB) in Fig. \ref{fig:s43} (a) can achieve a consensus since more than 50\% of nodes are successful, as a comparison, when SIR $= -6$ dB in Fig. \ref{fig:s43} (b), the consensus can not be achieved since more than 50\% nodes are invalid for transactions due to the poor communication channel quality. Analytical results are available in \cite{Xu2020a} for more details. Moreover, PBFT can be decomposed into multiple Raft procedures. For PoW, as we can see from Fig. 1, there is no need for uplink communication, hence the green circled area represents the PoW mining process.

\subsection{Throughput and Latency}
\begin{figure}[tbhp]
\centering
\includegraphics[width =0.8\textwidth]{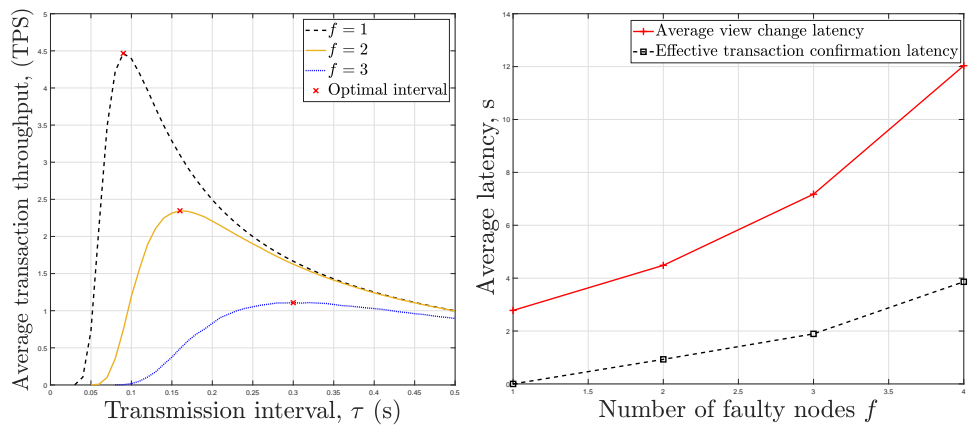}
\caption{Average transaction throughput (left) and latency (right) of wireless PBFT network}
\label{fig:V1}
\end{figure}

In the illustrative example in  Fig. \ref{fig:V1}, we show the impact of the transmission interval on the average transaction throughput and latency of the wireless PBFT network. In Fig. \ref{fig:V1} (a), we plot the average end-to-end average transaction throughput of the wireless PBFT network against the transmission interval for $n=10$ node and number of faulty nodes $f=1,2$ and $3$.  It can be seen that  there is an optimal transmission interval that maximizes the average end-to-end transaction throughput. This is due to the fact that the probability of a node to successfully receive  messages from other nodes increases as the transmission interval increases up to a given point. %Moreover, increasing the transmission interval beyond this point does not yield any further increase in the message reception success probability. Hence, increasing the transmission interval beyond the optimal value results in a decrease in the average transaction throughput. 
The effective use of the communication resource can thus be achieved with the utilization of the optimal transmission interval in the WBN design.  Fig. \ref{fig:V1} (a) further shows that increasing the number of faulty node $f$ increases the optimal transmission interval while reducing the optimal end-to-end transaction throughput. Fig. \ref{fig:V1} (b) shows the average transaction confirmation latency for the wireless PBFT with $n=15$ nodes by using the optimal transmission interval. It can be observed that increasing the number of faulty nodes $f$ leads to an increase in the transaction confirmation latency.
%\begin{figure}[tbhp]
%\includegraphics[width =0.5\textwidth]{FigV1.eps}
%\caption{End-to-end transaction throughput of wireless PBFT network with $n=10$ %nodes}
%\label{fig:V1}
%\end{figure}
%\begin{figure}[tbhp]
%\includegraphics[width =0.5\textwidth]{FigV2.eps}
%\caption{Average confirmation delay of the wireless PBFT network with $n=15$}
%\label{fig:V2}
%\end{figure}
\section{Communication protocol and topology}
Understanding the communication limits of wireless blockchain, other determining factors for the four metrics are communication protocols and typologies. It is well known that mesh-like topology is often regarded as the default topology for blockchain networks due to their heavy loads on information cross-check and data synchronization. Particularly, in the large mesh network, message relay is important for the coverage and performance issues regarding the Rx sensitivity and Tx power. As the mesh is the de facto topology, the protocols are the second determinant for the network characteristics, which mainly consists of two aspects: unicasting and broadcasting. 
\paragraph{Unicasting Protocol in Blockchain}
the term unicasting refers to a one-to-one transmission between the source and destination node. The unicast connection topology applies to most consensus algorithms. Examples of unicasting in blockchain include implementation of the client request procedure in PBFT and Raft consensus networks.
\paragraph{Broadcasting Protocol in Blockchain} the broadcasting protocol is required for the consensus and state replication phases of blockchain. In the consensus phase of PoW, the node that receives the client request broadcasts the transaction to other nodes in the consensus network, as shown in Fig. \ref{fig:s31} (b). Similarly, in the pre-prepare phase of PBFT, the primary node broadcasts the pre-prepare message to other nodes. Moreover, with successful pre-prepare, each node broadcasts a prepare message to other nodes, and subsequently broadcasts a commit message in the commit phase, as illustrated in Fig. 1 (c).
%\begin{figure*}
%\centering
%	\includegraphics[scale=0.65]{Delayvscpggplot2.pdf}
%	\caption{Delay vs Capture ratio of Flood Broadcast (a: left) and Unicast (b: right)}
%	\label{fig:s52}
%\end{figure*}
%\vspace{-5mm}
Broadcasting in wireless blockchain could be based on the conventional one-to-many transmission, where the source node sends a message to recipients within its coverage simultaneously. Alternatively, broadcasting could also be achieved by sending the same message one by one using the unicast connection between the sender and each recipient node within the sender's coverage. The sender creates multiple copies of the same message with different destination addresses. However, the approach suffers from significant delay, which scales with the number of nodes within the coverage of the sender/source node. It also consumes lots of bandwidth, and the sender must know the address of each of the destination nodes. 
Transaction broadcast is a common feature in most consensus algorithms. However, due to the limited propagation range of wireless transmission, the message broadcasted by the source node using the conventional broadcast or the unicast-based broadcast scheme might not reach all the nodes in the consensus network. Each of the recipients could re-broadcast the message using techniques such as flooding \cite{karp2000randomized} and gossip-based information propagation \cite{Williams02} to ensure that all nodes in the consensus network receive a copy of the message. Flooding starts in the wireless consensus network, with the primary node broadcasting a message to all its neighbor nodes \cite{karp2000randomized}. Afterward, every node that receives the message re-broadcasts it exactly once, and the process continues until all the reachable nodes in the network have received the message. 

\section{Conclusions} 
In this paper, we have analyzed the role of communication to the blockchain with three considered CMs, i.e., PoW, PBFT and Raft, and the impacts on the blockchain performance in terms of transaction throughput, latency, security and scalability. The paper considered different communication protocols and topology, such as unicast and broadcasting, with and without relay nodes. It has been shown that the communication resource is required in client requests, consensus achieving, and state replication and reply to the client. Analysis of communication complexity and provision for the three commonly used CMs shows that PBFT the most communication resource-demanding CM, while PoW is the least demanding one. In addition, Tx power and Rx sensitivity are discussed, where the simulation result shows that sufficient communication resource is essential to a secure wireless blockchain network. The paper provides useful guidance for a real wireless blockchain network deployment.

%%%%%%%%%%%%%%%%%%%%%%%%%%%%%%%%%%%%%%%%%%%%%%%%%%%%%%%%%%%%%%%%%%%%%
\normalem
\bibliographystyle{IEEEtran}
\bibliography{bcmag}
%%%%%%%%%%%%%%%%%%%%%%%%%%%%%%%%%%%%%%%%%%%%%%%%%%%%%%%%%%%%%%%%%%%%%

\end{document}